\documentclass{aastex631}

\shorttitle{Low-latitude Sunspots Group}
\shortauthors{Tlatov }
\graphicspath{{./}{figures/}}

\begin{document}

\title{ Low-Latitude Sunspot Group Tilt Angles in 15\,-\,24 Activity Cycles
}

\author[0000-0002-6286-3544]{Andrey G. Tlatov}
\affiliation{Kislovodsk Mountain Astronomical Station of the Pulkovo observatory, Kislovodsk, Gagarina str. 100, 357700, Russia}
\affiliation{Kalmyk State University, Elista, Russia}

\begin{abstract}
An analysis of the tilt angles of the active regions in 15\,-\,24 activity cycles was performed.  We used data from measurements of magnetic fields in the sunspot umbra in the period 1918\, -\,2019 at the Mount Wilson Observatory, as well as the tilt angles of active regions in 'white' light at the Kodaikanal and Mount Wilson observatories in activity cycles 15\, -\,21.
The mean tilt angles of active regions $\overline{\gamma}$  and the slope $\mu$ from latitude $\theta$ in the activity cycles are considered.
Low-latitude bipoles are the most important in predicting the strength of solar cycles. In this work, we selected the cutoff latitude $\theta_{cut}$ at which the highest correlation is observed with the strength of the next activity cycle for active regions with latitude $\theta<\theta_{cut}$. It was found that for magnetic field measurement data, the highest correlation of the parameters
$\overline{\gamma}$ and $\mu$ with the strength of the next solar activity cycle is characteristic of bipoles in the equatorial zone with $\theta<\theta_{cut}\approx 14.2^o$. 
For ‘white’ light observation data, $\theta_{cut}\approx 8.5^o$ for Mount Wilson observatory and $\theta_{cut}\approx 9.4^o$ for Kodaikanal observatory.

\end{abstract}


\section{Introduction} \label{sec:intro}

The direction of the magnetic axis of the solar bipoles connecting the leading and trailing parts has a distinct angle $\gamma$ (tilt angle) with the east-west line direction. As a rule, the trailing parts of sunspot groups or active regions (AR) are closer to the poles, and the tilt angle increases with latitude. This pattern was established by Joy \citep{Hale1919}. Later, Joy's law was confirmed in many studies \citep{Howard1991, Sivaraman.eta1993}. It is believed that this law, together with Hale's law \citep{Hale1919} on the change in polarity of the leading and trailing sunspots in a 22-year cycle, is an important reason for the existence of the solar magnetic cycle \citep{Babcock1961, Leighton1969}. The tilt angle is the parameter that links the toroidal magnetic fields of the current cycle with the activity of the next cycle.

The tilt angle measurements are made from different data. \cite{Howard.etal1984}   made the tilt angle measurements from observations of active regions in white light using  Mount Wilson Observatory (MWO) data in the period 1917\,-\,1985 and \cite{Howard.etal1999}  the Kodaikanal Observatory (KK) 1906\,-\,1987. The KK data cover activity cycles 14\,-\,21, but the KK data are missing 4 years of cycle 14. The MWO data are missing the first 4 years of cycle 15 and one year of cycle 21. These data did not use magnetic polarity data \citep{Howard1991}.

From these data, it was found that the tilt angles and Joy's law depend on several parameters of sunspot groups, including latitude, area, formation and evolution, rotation and meridional motions, and the phase of the solar cycle \citep{Howard1991,Sivaraman.etal1999}. \cite{Dasi-Espuig.etal2010} analyzed tilt angle data from MWO and KK and found relationships between the tilt angle and the solar cycle strength.  \cite{Ivanov2012} repeated the analysis of KK and MWO observations, including the Pulkovo/Kislovodsk sunspot group catalog, covering the period of cycles 18\,–\,21. Other active region tilt databases are also used. \cite{Baranyi2015}  presented databases based on the Debrecen sunspot catalogs. \cite{Isıket.al2018} measured the tilt angle from solar drawings made at the Kandilli Observatory for cycles 19\,–\,24. The tilt angles of active regions can be reconstructed from the beginning of sunspot observations and drawings in the early 17th century \citep{Clette2023, Hayakawa2021, Hayakawa2024, Muñoz-Jaramillo}.

The data sets discussed above were based on sunspot positions, without information on magnetic polarity. Active region tilts, taking into account the magnetic polarity of the sunspots, were determined by \cite{Li.Ulrich2012} from Mount Wilson and MDI magnetograms. \cite{McClintock2014}  compared these measurements with the Debrecen tilt data, focusing on the anti-Hale regions, which are the main cause of the discrepancies. The magnetogram-based data give systematically larger values. \cite{Howard1996b}  obtained an average angle $\overline{\gamma}\approx 6.3^{o}$ based on magnetic field observations in the period 1967\,–\,1995. This is consistent with the data of \citep{Wang.etal2015}. The magnetographic observation data start from activity cycle 21.

Active region tilts taking into account the magnetic polarity of sunspots were also determined by \cite{Tlatova.etal2018}  from the archive of solar drawings at Mount Wilson Observatory. These drawings include information on the polarity and intensity of the magnetic field in sunspots umbra  since 1918. The data of \citep{Tlatova.etal2018}, obtained by Gaussian fitting for solar cycles 15\,-\,24, did not confirm the relationship between the average tilt angle $\overline{\gamma}$ and the amplitude of the current or subsequent cycle found in \citep{Dasi-Espuig.etal2010}.

Joy's law can be explained by the action of the Coriolis force, which rotates the emerging magnetic fields that form sunspots \citep{Fisher.etal1995}. The orientation of the bipoles is close to random during the first appearance of the bipoles, but is ordered after 1 to 3 days \citep{Weart, Harvey1993}. \cite{Harvey1993} was the first to show that the orientation of the bipoles depends on the size of the leading and trailing parts of the bipoles (i.e., on the magnetic flux).

\begin{figure}[ht!]
\centerline{\includegraphics[width=0.8\textwidth,clip=]{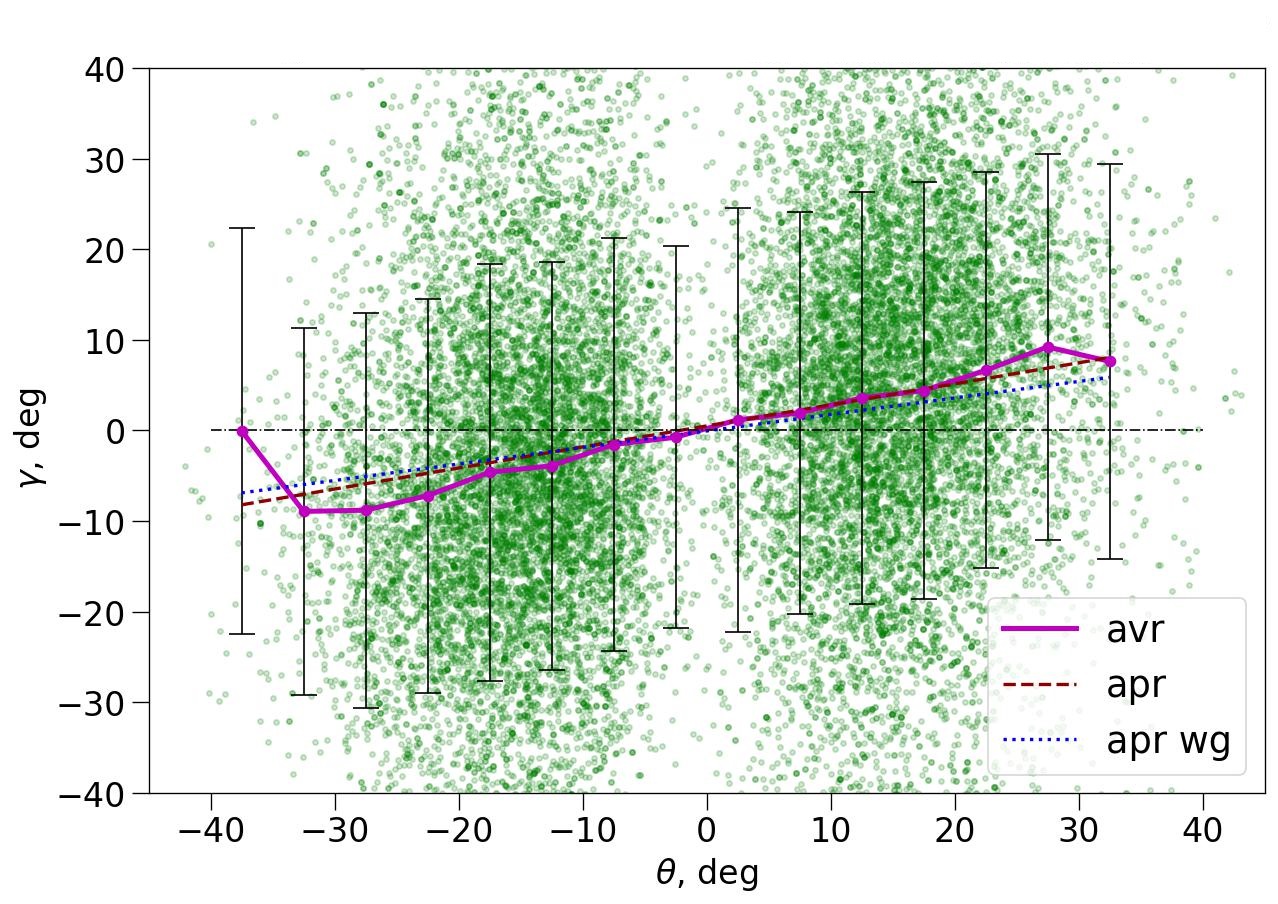}
                 } 
\caption{Magnetic bipoles tilt angles $\gamma$ from latitude $\theta$ based on observations of sunspot magnetic fields by  Mount Wilson observatory in the period 1918\,-\,2019. The solid line shows the average values in 5-degree intervals by latitude $\theta$, as well as the standard deviations. The dashed line shows the linear approximation from latitude. The dashed lines show an area-weighted approximation.}
\label{fig:fig1}
\end{figure}

\begin{table}
\centering
\tabletypesize{\scriptsize}
\tablewidth{700pt} 
\caption{Parameters of tilt angles in solar cycles. $[N_{cl}]$ are the activity cycle number; [$k$] are the total number of bipoles; [$kN$] are the number of bipoles in the northern and [$kS$] southern hemispheres; values of mean tilt angle [$\overline{\gamma}$] in degree and the  standard deviations [$\sigma$]; values of the intercept [b] and the slope [$\mu$] and the corresponding standard deviations [$\sigma_{b}$]  and [$\sigma_{\mu}$],  calculated without and with  weight [$b_w$], [$\mu_w$]. }

\label{T1}


\begin{tabular}{rrrrrrrrrrrrrr} 
 \hline 
 $N_{cl}$ &   k   &   kN  &   kS  &    $\overline{\gamma}$  &    $\sigma$  &   $\mu$   &  $\sigma_{\mu}$  &   b  &   $\sigma_{b}$   &   $\mu_{w}$ &  $\sigma_{\mu w}$  &  $b_{w}$   &  $\sigma_{bw}$\\ 
 \hline
 {MWOMF} \\
\hline 
  15 &  2031 &   985 &  1046 &     4.61 &    1.03 &     0.49 &  0.045 &   -0.75 &    0.59 &     0.50 &  0.044 &   -2.22 &    0.59 \\ 
 16 &  4638 &  2436 &  2202 &     4.26 &    0.56 &     0.43 &  0.023 &   -1.07 &    0.37 &     0.29 &  0.022 &   -2.74 &    0.35\\  
 17 &  2914 &  1415 &  1499 &     3.20 &    0.69 &     0.24 &  0.026 &    0.36 &    0.47 &     0.20 &  0.025 &    0.74 &    0.43 \\ 
 18 &  3357 &  1698 &  1659 &     6.64 &    0.79 &     0.32 &  0.028 &    0.90 &    0.45 &     0.41 &  0.025 &    1.72 &    0.41 \\ 
 19 &  3994 &  2295 &  1699 &     5.08 &    0.61 &     0.26 &  0.022 &   -0.05 &    0.43 &     0.30 &  0.020 &   -0.56 &    0.39 \\ 
 20 &  7723 &  4361 &  3362 &     4.34 &    0.53 &     0.28 &  0.020 &   -0.32 &    0.31 &     0.29 &  0.019 &    0.80 &    0.30 \\ 
 21 &  8965 &  4334 &  4631 &     2.91 &    0.58 &     0.27 &  0.019 &   -0.65 &    0.31 &     0.19 &  0.017 &   -0.93 &    0.29 \\ 
 22 & 11426 &  5355 &  6071 &     2.76 &    0.55 &     0.25 &  0.015 &   -0.34 &    0.28 &     0.19 &  0.015 &   -1.63 &    0.25 \\ 
 23 &  7088 &  3525 &  3563 &     2.79 &    0.63 &     0.26 &  0.018 &    0.80 &    0.33 &     0.21 &  0.017 &   -0.34 &    0.30 \\ 
 24 &  3469 &  1548 &  1921 &     2.77 &    0.73 &     0.30 &  0.026 &    0.44 &    0.42 &     0.22 &  0.026 &    0.43 &    0.41 \\ 
 All& 55605 & 27952 & 27653 &     3.83 &    0.83 &     0.28 &  0.007 &   -0.13 &    0.12 &     0.25 &  0.006 &   -0.49 &    0.11  \\ 
\hline 
{KK} \\
\hline 
  14 &  1611 &   746 &   865 &     3.09 &    0.98 &     0.26 &  0.061 &    1.30 &    0.76 &     0.20 &  0.057 &    1.24 &    0.73\\  
 15 &  2800 &  1475 &  1325 &     4.87 &    0.66 &     0.36 &  0.039 &    1.19 &    0.58 &     0.37 &  0.036 &    0.71 &    0.57\\  
 16 &  3350 &  1747 &  1603 &     5.90 &    0.55 &     0.33 &  0.032 &    0.90 &    0.51 &     0.35 &  0.030 &   -0.22 &    0.50 \\ 
 17 &  4576 &  2252 &  2324 &     6.39 &    0.54 &     0.29 &  0.028 &    1.01 &    0.46 &     0.39 &  0.025 &    1.52 &    0.45 \\ 
 18 &  4451 &  2248 &  2203 &     4.85 &    0.54 &     0.23 &  0.028 &    0.50 &    0.45 &     0.26 &  0.026 &    1.26 &    0.44 \\ 
 19 &  4900 &  2839 &  2061 &     4.41 &    0.49 &     0.22 &  0.024 &    1.20 &    0.45 &     0.22 &  0.024 &    1.77 &    0.42 \\ 
 20 &  4956 &  2718 &  2238 &     5.66 &    0.47 &     0.26 &  0.026 &   -0.48 &    0.43 &     0.34 &  0.026 &    0.13 &    0.39 \\ 
 21 &  3882 &  1934 &  1948 &     5.22 &    0.54 &     0.23 &  0.028 &    0.57 &    0.50 &     0.29 &  0.026 &    1.32 &    0.49 \\ 
  All& 30614 & 15989 & 14625 &     5.26 &    3.68 &     0.26 &  0.011 &    0.66 &    0.18 &     0.31 &  0.010 &    0.97 &    0.16 \\ 
\hline 
{MWO} \\
\hline
 15 &  2253 &  1142 &  1111 &     5.56 &    0.74 &     0.33 &  0.049 &    0.50 &    0.67 &     0.43 &  0.045 &    0.06 &    0.66 \\ 
 16 &  3080 &  1641 &  1439 &     4.95 &    0.60 &     0.29 &  0.035 &   -0.34 &    0.54 &     0.33 &  0.034 &   -0.92 &    0.49\\  
 17 &  3799 &  1892 &  1907 &     5.72 &    0.55 &     0.29 &  0.031 &    0.56 &    0.50 &     0.36 &  0.030 &    1.67 &    0.49\\  
 18 &  4952 &  2391 &  2561 &     5.65 &    0.47 &     0.26 &  0.026 &    1.29 &    0.43 &     0.33 &  0.026 &    1.30 &    0.40\\  
 19 &  6009 &  3470 &  2539 &     3.80 &    0.41 &     0.18 &  0.022 &    1.23 &    0.40 &     0.18 &  0.020 &    1.75 &    0.39 \\ 
 20 &  4434 &  2438 &  1996 &     4.53 &    0.49 &     0.28 &  0.029 &   -0.73 &    0.45 &     0.33 &  0.027 &   -0.12 &    0.41\\  
 21 &  3828 &  1934 &  1894 &     5.20 &    0.52 &     0.29 &  0.028 &    0.27 &    0.48 &     0.31 &  0.028 &    0.84 &    0.45\\  
 All& 28355 & 14908 & 13447 &     4.91 &    0.53 &     0.26 &  0.011 &    0.44 &    0.18 &     0.29 &  0.011 &    0.77 &    0.18 \\ 
\hline
\end{tabular} 
\end{table}

\section{Data} \label{sec:data}

One of the data sets in the work is a digitization of measurements in the magnetic field of the sunspot umbra of the Mount Wilson Observatory (MWOMF)  for the period 1918\,-\,2019 (\url{https://www.mtwilson.edu/solar-observing/}). These observations are carried out on the tower spectrograph by visually measuring the splitting of the Zeeman component in the iron line $\lambda6173 \AA$ in the period 1917\,-\,1962, and after this period in the line $\lambda 5250 \AA$.  The drawings were digitized using a dedicated software package \citep{Tlatova.etal2018}. The digitization included the date and time of observations, the heliographic coordinates of each umbra, its area, and the strength and polarity of its magnetic field. The digitized data sets used in this paper contains 20,318 days of observations. The total number of umbrae and pores in these images is about $5\,10^5$. The first results of the analysis of the data set were described in \citep{Tlatova.etal2018}.

One of the clear advantages of MWOMF sunspot patterns over previous tilt angle studies using historical data is that knowledge of sunspot polarities allows better identification of sunspot pairs that form a group. Based on the magnetic field strength, position and area of the measured sunspots, a database of magnetic bipoles was created for the period of the 15th-24th activity cycles. The 15th activity cycle is not fully represented in the MWOMF data.

The algorithm is based on a sample of nearby sunspots of opposite magnetic field polarity. For the selected bipoles, the tilt angle of the magnetic axis to the equator was determined. The axis was drawn from the center of the leading (western) sunspot to the center of the trailing (eastern) sunspot. To calculate the tilt angles, we use the formula \citep{Howard1991} $tan\gamma=\Delta \theta/\Delta \phi cos\theta$, where $\theta$ is the heliographic latitude, $\Delta \theta$ is the latitudinal and $\Delta \phi$ is the longitudinal angular separation, respectively. As magnetic centers, we used the geometric mean for clusters of sunspots of leading and trailing polarity. Centroids of positive and negative polarities are calculated using flux weighting. This method is a traditional and widely accepted approach to determine the tilt angle \citep{Howard1991}.

From the digitization itself, we can take only the areas of the sunspot umbra. In this work we also compared the found bipoles of active regions with the area of the sunspot groups $S_{gr}$ presented on the website \url{http://solarcyclescience.com/activeregions.html} (RGO-USAF). For this, on the minimum time interval for the bipoles identified in the MWOMF data, the parameters of the RGO-USAF sunspot groups were found, with an area greater than the total area of the umbrae in the sketches and located at a minimum distance from the coordinates of the group. In total, we identified 57214 sunspot groups with measured magnetic axis directions. The average area of such groups was $S_{gr}\approx 240.8$ millionths of the solar hemisphere ($\mu$sh). In order to reduce the error, we further limited the groups by the distance from the central meridian $\phi< \pm 70^{o}$. In total, there were 55520 such groups. In the following, we used the area of the groups and coordinates taken from the RGO-USAF white-light sunspot processing data for the selection.

Other data were active region digitization data from Mount Wilson white light observations in the period 1917\,–\,1985 \citep{Howard.etal1984} and Kodaikanal 1906–1987 \citep{Sivaraman.eta1993}
(\url{www.ngdc.noaa.gov/stp/solar/sunspotregionsdata.html}). The technique for finding the tilt angles $\gamma$ is based on measuring the position and area of individual sunspots. The sunspot data were grouped into active region groups using the proximity-based method of \citep{Howard.etal1984}. To distinguish between leading and trailing parts of sunspot groups, the authors first calculated the center of mass. The part to the east of the center of mass was defined as the leading part, and the part to the west as the trailing part. The KK and MWO data sets are widely used in active region tilt angle analysis  (see \cite{Petrovay2020}).

In an analysis by \cite{Dasi-Espuig.etal2010}, it was suggested that for active regions close to the equator, the magnetic flux of the leading parts of the active regions can cross the equator and cancel the opposite polarity of the leading part of the active region in the other hemisphere. Consequently, active regions at low latitudes contribute disproportionately to the reversal and accumulation of the magnetic field at the poles. This is thought to influence the strength of the next cycle, since the polar fields are the input to the next cycle. \cite{Dasi-Espuig.etal2010} accounted for this latitude dependence by multiplying an exponential function of latitude by area-weighted tilt angles. The correlation coefficients between these parameters and the amplitude of the next activity cycle reach $r=0.79$ for MWO and $r=0.78$ for KK.

\begin{figure}
 \includegraphics[width=0.48\textwidth,clip=]{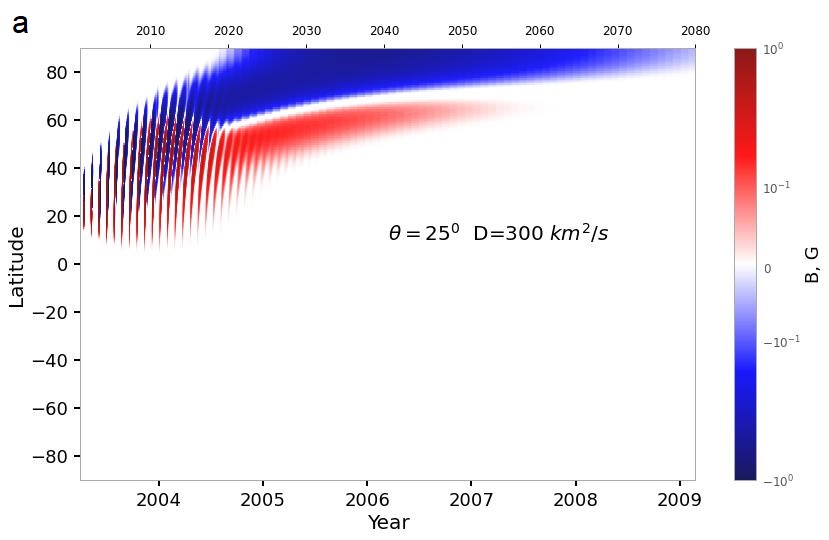}
               \hspace*{0.013\textwidth}
               \includegraphics[width=0.48\textwidth,clip=]{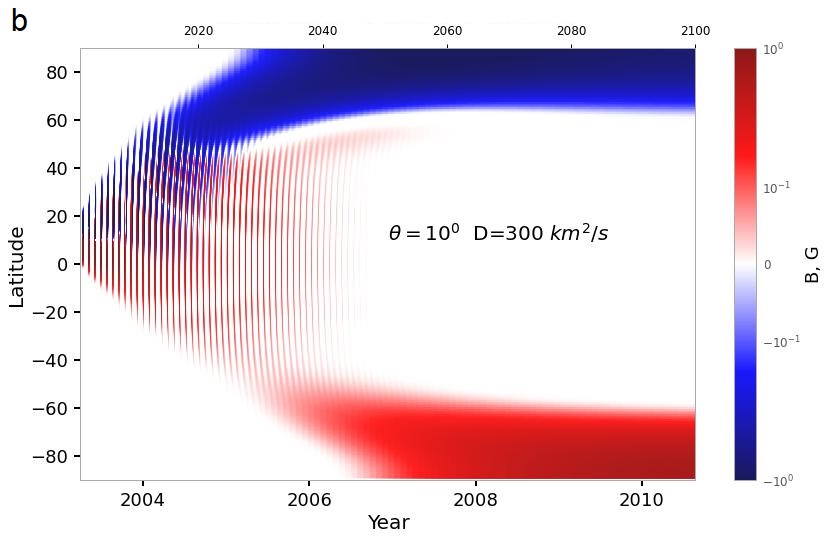}
\caption{Supersynoptic map of the magnetic field in latitude-time coordinates for a test bipole initially placed at latitude: a) $\theta1 =25^o$ with a diffusion coefficient $D = 300 km^{2}\,s^{-1}$; b) same as (a), but for a bipole at latitude $\theta2 =10^o$.
}
\label{fig:fig2}
\end{figure}

\section{Tilt Angles of Active Regions}
\subsection{Latitude dependence of magnetic bipoles tilt angles}
One of the observational data in this work was the Mount Wilson Observatory magnetic field observations. One distinct advantage of the MWOMF sunspot patterns over all previous tilt angle studies using historical data is that knowledge of sunspot polarities allows better identification of sunspot pairs that form the sunspot group.
The dependence of tilt angles on latitude $\theta$ can be investigated using various parameters. These can be average values $\overline{\gamma}$ in latitude intervals taken with weight by area of AR, or without taking weight into account. The search for maxima of the Gaussian distribution is also used \citep{Tlatova.etal2018}. Therefore, the results of studies may vary.

Another parameter for describing the slope is the parameter $\mu$ in the linear approximation $\gamma=b+\mu\theta$, where $\theta$ is the latitude in degrees. The calculation of parameters $b$ and $\mu$ can be performed with and without area weighting. In the approximation without weighting, the residual sum of squares is minimized: $RSS(\mu)=\Sigma(y_{i}-x_{i}\mu)^{2}$. With weighting, we minimize the weighted sum of squares $WSS(\mu)=\Sigma \omega_{i}(y_{i}-x_{i}\mu)^2$. Area weighting reveals clearer patterns \citep{Javaraiah2023}.

Figure \ref{fig:fig1} shows the scatter plot of the tilt angles of the MWOMF magnetic bipoles in the period 1918,-,2019. The plot has a fairly large spread of tilt angle values at the latitudes of sunspot existence. To illustrate this, Figure \ref{fig:fig1} shows the average values in 5-degree latitude intervals calculated taking into account the weight by the area of groups and the magnitude of the square deviation $\sigma$. The dotted line shows the approximation without taking into account the weight $\gamma=b+\mu\theta=0.5(\pm0.07)+0.23(\pm0.01)\theta$ and taking into account the weight $\gamma=b_{w}+\mu_{w}\theta=-0.061(\pm0.085)+0.182(\pm0.005)\theta$. We used the area of sunspot groups $S_{gr}$ as the weight.

\subsection{Tilt Angles of Active Regions in Solar Cycles}

Table \ref{T1} presents the parameters of the tilt angles of active regions, including the average tilt angle $\overline{\gamma}$ calculated using the AR area weight, the slope parameter $\mu$ and $\mu_{w}$ calculated without and using the weight.
To calculate the average values from the MWOMF data, we used bipoles with tilt angles $\gamma$ in the range $80^{o} >abs(\gamma)>0$.

The highest average tilt angle according to MWOMF was in the activity cycle 18 $\overline{\gamma}\approx 6.64^{o}$,  according to KK in the  activity cycle 17 $\overline{\gamma}\approx 6.39^{o}$, and according to MWO in the  activity cycle 18 $\overline{\gamma}\approx5.65^{o}$. 

\cite{Dasi-Espuig.etal2010} analyzed MWO and KK tilt angle data for solar cycles 15\,–\,21. Area-weighted average tilt angles of sunspot groups were calculated in latitude bins and then normalized by latitude to obtain the tilt parameter. A fairly good anti-correlation was found between the ratio of the mean tilt angle $\overline{\gamma}$ over a solar cycle to the mean absolute latitude $\overline{\theta}$ over the same solar cycle and the amplitude of the solar cycle (SN).  Subsequent studies using different data sets once questioned its existence \citep{Ivanov2012, Tlatova.etal2018, Isıket.al2018}. However, using latitude normalization may give an incorrect interpretation of the relationship, since the average latitude of AR has a correlation with the amplitude of the activity cycle \cite{Jiang2020}. The $\overline{\gamma}$ values in Table \ref{T1} for the KK and MWO data are close to the results of \citep{Dasi-Espuig.etal2010}, with small deviations due to different data filtering.
According to the values of the weighted tilt angles found by us, there is no correlation between $\overline{\gamma}$ and the amplitude of the cycle. There is also no relationship between the parameter $\mu_{w}$ and the amplitude of the current activity cycle ($SN$) according to the MWOMF and KK data. But according to MWO data, the correlation between$\mu_{w}$ and $SN$ is $r=-0.79$.

\subsection{Tilt angle of Active Regions and the Strength of the Next Cycle of Activity}

Several authors  \citep{Sivaraman.etal1999, Dasi-Espuig.etal2010},  reported that for cycles 16\,–\,21, the latitude-normalized mean tilt angles correlate with the amplitude of the next solar cycle. However, later studies \citep{Ivanov2012,McClintock.Norton2013} failed to reproduce the results of \cite{Dasi-Espuig.etal2010}, and later \citep{Dasi-Espuig.etal2013} revised their earlier results. \cite{Javaraiah2023} found that there is a significant anti-correlation between the MWO tilt angle-derived data and the solar cycle amplitude for the southern hemisphere, while there is no significant correlation for the northern hemisphere.

The first sunspots of a new activity cycle appear at mid-latitudes, but as the cycle progresses, the mean latitude of their appearance shifts toward the equator. This feature of the sunspot cycle development is called Spörer's law. The speed of drift toward the equator is greater at the beginning of the cycle and slower at the end of the cycle \citep{Hathaway2015}. \cite{Vitinskij} used sunspot latitudes near minimum as a predictor of cycle amplitude in 21. Sunspot latitude is now considered a poor indicator of cycle amplitude \citep{Hathaway2015}.

Previous studies show that low-latitude ARs can significantly affect the evolution of the polar field \citep{Cameron.etal2013,Jiang2015,Nagy.eta2017}. As is known, AR appear at lower latitudes during the declining phase. Therefore, compared with AR during the ascending and maximum phases, which appear at higher latitudes, AR during the declining phase should have a greater influence on the evolution of the polar fields. The low-latitude magnetic flux is transported poleward, taking about 2 years. Thus, the AR tilt angles during the decay phase (after the polarity reversal) are more important for the build-up of the polar field of the opposite polarity, which is the source. \cite{Gao} found that the latitude-weighted absolute value of the tilt angles during the decay phase is significantly anti-correlated with the strength and amplitude of the next solar cycle. This contradicts the results of previous studies that there should be a positive relationship between the AR tilt angles at low latitudes and the evolution of polar fields \citep{Petrovay2020}.

\begin{table}
\centering
\tabletypesize{\scriptsize}
\tablewidth{700pt} 
\caption{Parameters of  tilt angles of  low-latitude  ARs $\theta<\theta_{cut}$. $[N_{cl}]$ are the activity cycle number; [$k$] are the number of ARs; [$\overline{\gamma^{\theta}}$] are the average tilt angle; [$\overline{\gamma^{\theta}sin\theta}$]   are the same but taking into account the sine of the latitude; [$\sigma$] are the standard deviations; [$\Sigma S^{\theta}$] are the sum of areas of AR; [$\mu$] are area-weighted slope approximation parameter;  [$\sigma_{\mu}$] are the standard deviations.}

\label{T2}
 
\begin{tabular}{crccccccccccccc}    
 \hline 
$N_{cl}$   &   k   & $\Sigma S^\theta$ & $\overline{\gamma^\theta} $   & $\sigma_{\gamma}$   & $\overline{\gamma^\theta \cdot sin(\theta)} $     &  $\mu^{\theta}_{w}$     &  $\sigma_{\mu}$ \\   
\hline
 {MWOMF \,\,\,   $\theta<14.2^{0}$} \\
\hline
 15 & 1311 &  1.32e+06 &     3.18 &    0.972 &       0.82 &    0.565 &    0.077\\  16 &  2285 &  1.25e+06 &     2.75 &    0.604 &       0.51 &    0.322 &    0.051\\  
 17 &  1302 &  1.90e+06 &     2.23 &    0.648 &       0.43 &    0.221 &    0.069\\  
 18 &  1688 &  2.15e+06 &     4.54 &    0.980 &       0.85 &    0.441 &    0.062\\  
 19 &  1314 &  2.05e+06 &     0.29 &    0.906 &       0.17 &    0.178 &    0.069\\  
 20 &  4173 &  1.85e+06 &     2.48 &    0.560 &       0.52 &    0.270 &    0.038\\  
 21 &  4356 &  2.40e+06 &     3.02 &    0.496 &       0.44 &    0.218 &    0.040\\  
 22 &  4374 &  1.66e+06 &     1.05 &    0.590 &       0.27 &    0.154 &    0.043\\  
 23 &  2302 &  1.86e+06 &     0.44 &    0.800 &       0.09 &    0.060 &    0.055\\  
 24 &  1766 &  1.03e+06 &     1.14 &    0.648 &       0.28 &    0.163 &    0.051\\  

 \hline
{KK \,\,\,   $\theta<9.4^{0}$} \\
\hline
 14 &   612 &  5.99e+05 &     2.95 &    0.114 &       0.26 &    0.341 &    0.173\\  
 15 &   836 &  7.17e+05 &     0.32 &    0.104 &       0.03 &    0.053 &    0.151\\  
 16 &   892 &  6.60e+05 &     2.69 &    0.089 &       0.31 &    0.252 &    0.131 \\ 
 17 &  1033 &  7.93e+05 &     5.64 &    0.099 &       0.58 &    0.661 &    0.139 \\ 
 18 &  1012 &  8.61e+05 &     6.24 &    0.094 &       0.78 &    0.956 &    0.124\\  
 19 &   943 &  8.53e+05 &     0.77 &    0.098 &       0.13 &    0.095 &    0.145\\  
 20 &  1139 &  8.06e+05 &     3.33 &    0.090 &       0.38 &    0.443 &    0.122\\  
 21 &   740 &  9.23e+05 &     4.77 &    0.088 &       0.51 &    0.498 &    0.161\\  

\hline 
{MWO \,\,\,   $\theta<8.5^{0}$} \\
\hline
 15 &   562 &  5.98e+05 &     1.48 &    1.458 &       0.12 &    0.186 &    0.221\\  
 16 &   704 &  5.29e+05 &     2.88 &    1.080 &       0.30 &    0.407 &    0.162\\  
 17 &   762 &  6.20e+05 &     3.35 &    1.045 &       0.24 &    0.407 &    0.175\\  
 18 &   877 &  6.49e+05 &     3.58 &    1.083 &       0.43 &    0.657 &    0.163\\  
 19 &   948 &  6.71e+05 &     2.22 &    0.962 &       0.24 &    0.359 &    0.157\\  
 20 &   924 &  6.45e+05 &     2.33 &    0.887 &       0.26 &    0.340 &    0.137\\  
 21 &   628 &  7.31e+05 &     3.16 &    1.304 &       0.19 &    0.321 &    0.201 \\

\hline
\end{tabular} 
\end{table}

\subsubsection{SFT Model}
The tilt angle of solar active regions is of key importance for the formation of the polar field. According to dynamo models of the Babcock-Leighton type, the large-scale magnetic field is formed from active regions of the bipolar type evolving under the action of diffusion, differential rotation $v_{\phi}$ and meridional circulation $v_{m}$. Given the dependence of $v_{m}$ and the tilt angles $\gamma$ on latitude, the contribution of active regions to the formation of the poloidal field will also be different.  We have developed software for Surface Flux Transport (SFT) modeling \citep{Yeates.etal2015, Tlatov.etal2024a}. Figure \ref{fig:fig2} shows the latitude-time plots for the test bipoles at latitudes $\theta=10^{o}$ and $\theta=25^{o}$ in the northern hemisphere. The bipoles were placed at longitude $180^o$ for the Carrington rotation 2000 (5 March 2003). The bipoles were tilted $\gamma=10^o$. The net magnetic flux was zero. For these test calculations, we used the SFT model with the diffusion coefficient D=300 $km^2\, s^{-1}$ and the meridional circulation velocity with the amplitude $v_{m}\approx10$ m $s^{-1}$ in mid-latitudes. To construct the diagram, we did not average by longitude but used data near the central meridian. 

The magnetic field at the poles for the bipole $\theta1$ tends to 0 over time. However, for the bipole $\theta2$ it becomes non-zero and forms the dipole-type magnetic field configuration. For the bipole $\theta2$, the polar magnetic field is formed over 2 years in the northern hemisphere and $\approx3.5$ years in the southern hemisphere. From test calculations it follows that for the formation of the polar magnetic field and the next cycle of activity, bipoles in the equatorial zone, limited by a certain latitude $\theta_{cut}$, are important. To predict solar cyclicity, not all bipoles of the current cycle can be considered, but only near-equatorial bipoles \citep{Yeates}.
It was found in \citep{Tlatov.etal2024b} that there is a significant correlation between active regions at low latitudes and the amplitude of the next activity cycle.

\begin{figure}
 \includegraphics[width=0.48\textwidth,clip=]{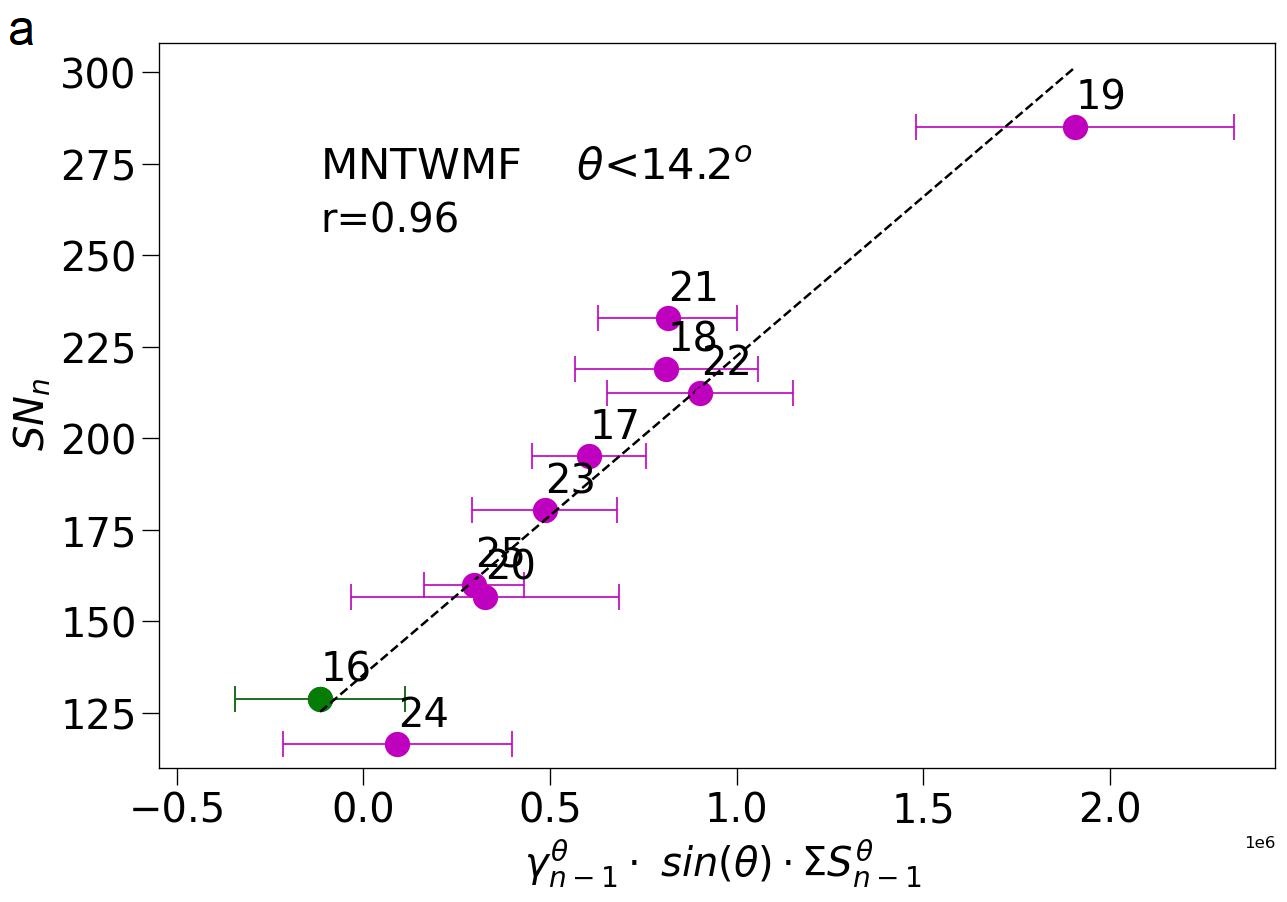}
               \hspace*{0.013\textwidth}
               \includegraphics[width=0.48\textwidth,clip=]{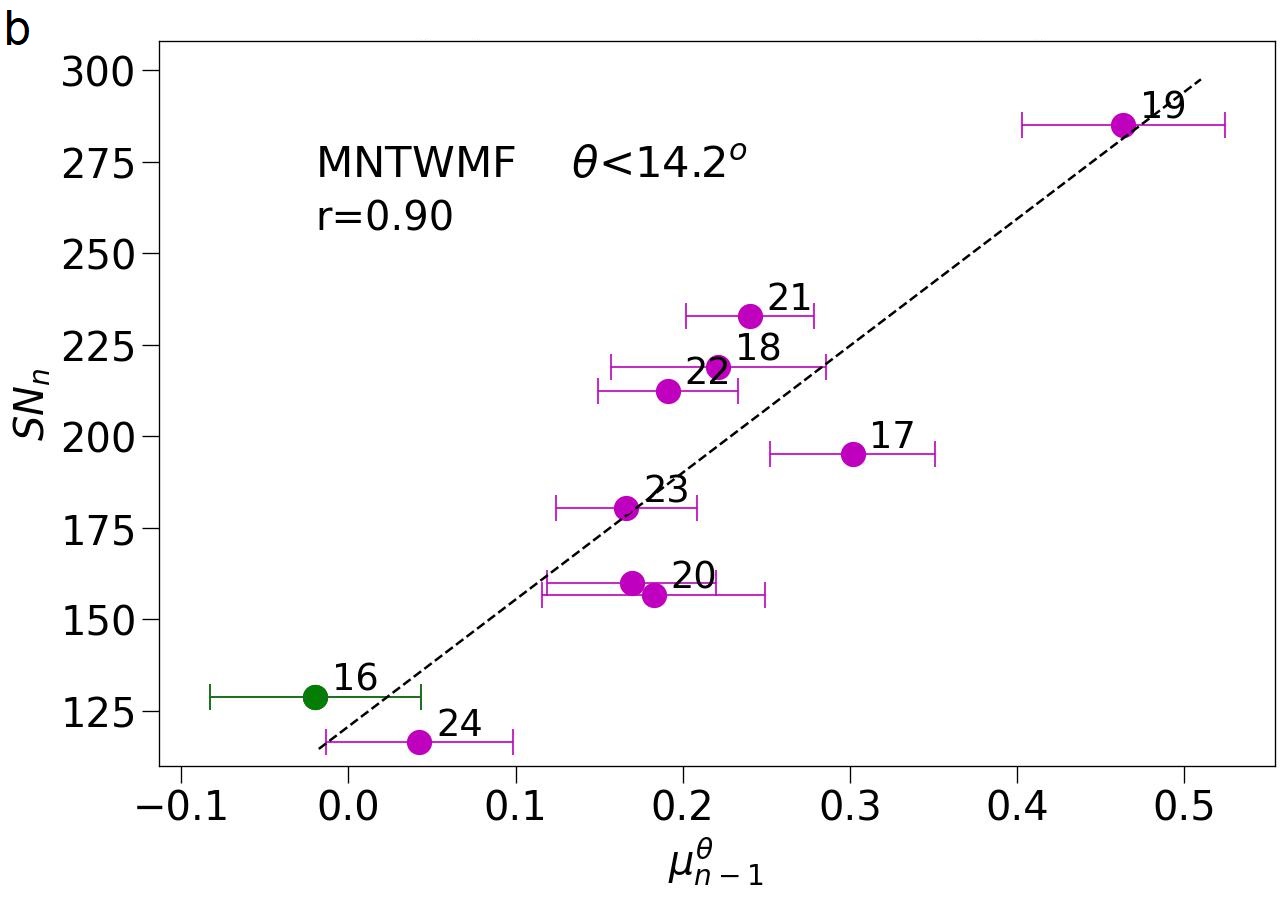}
 \includegraphics[width=0.48\textwidth,clip=]{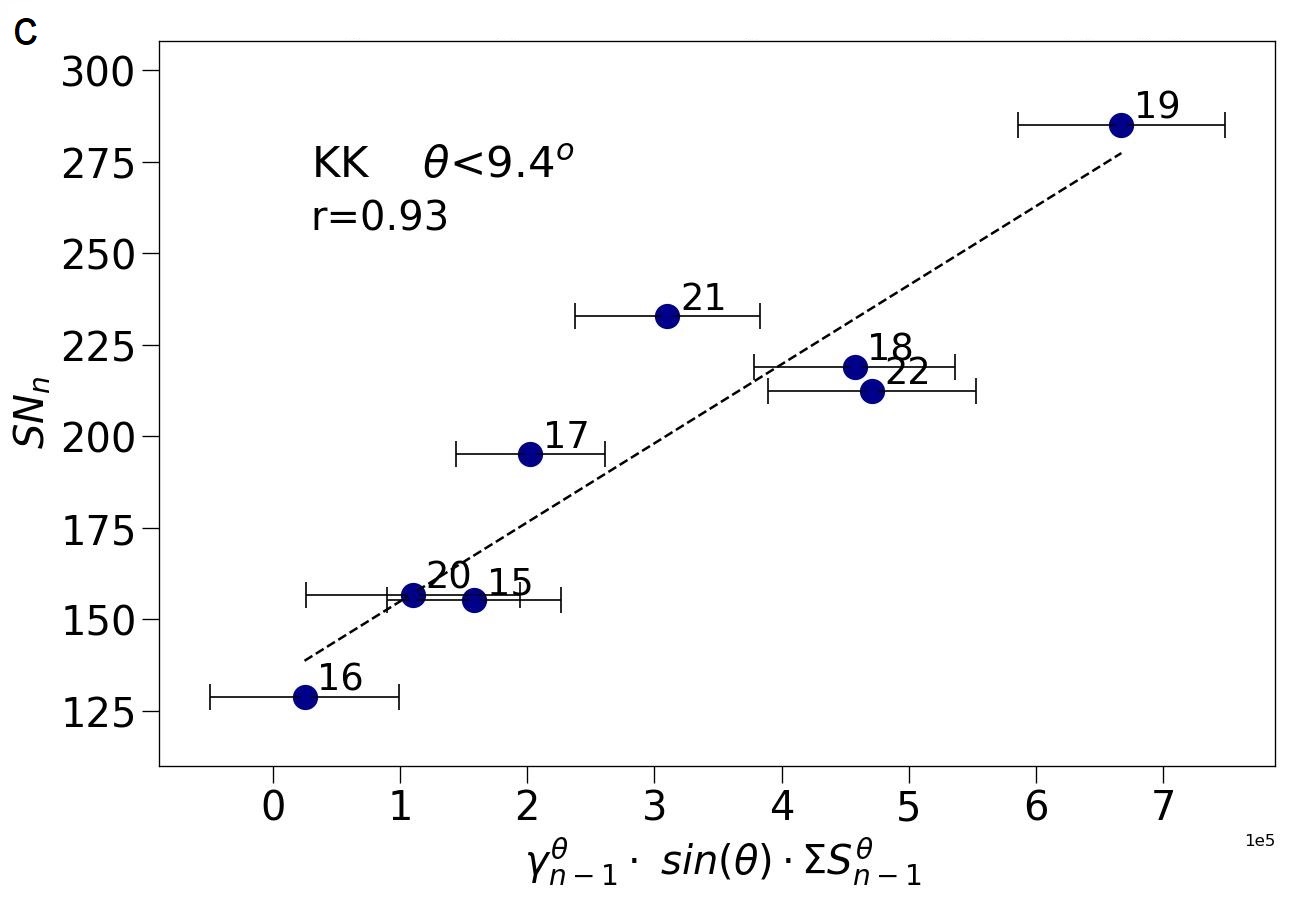}
               \hspace*{0.013\textwidth}
               \includegraphics[width=0.48\textwidth,clip=]{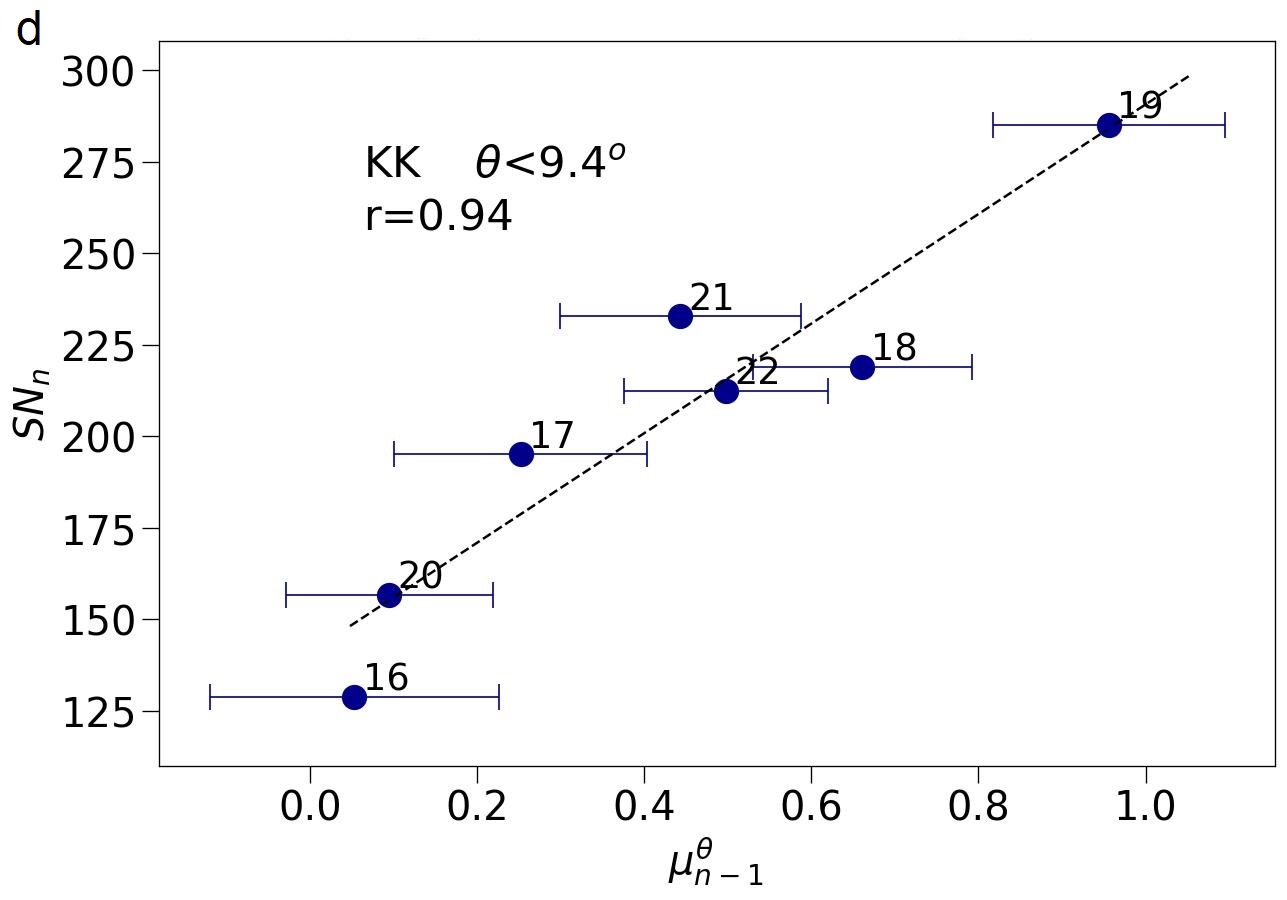}    
\includegraphics[width=0.49\textwidth,clip=]{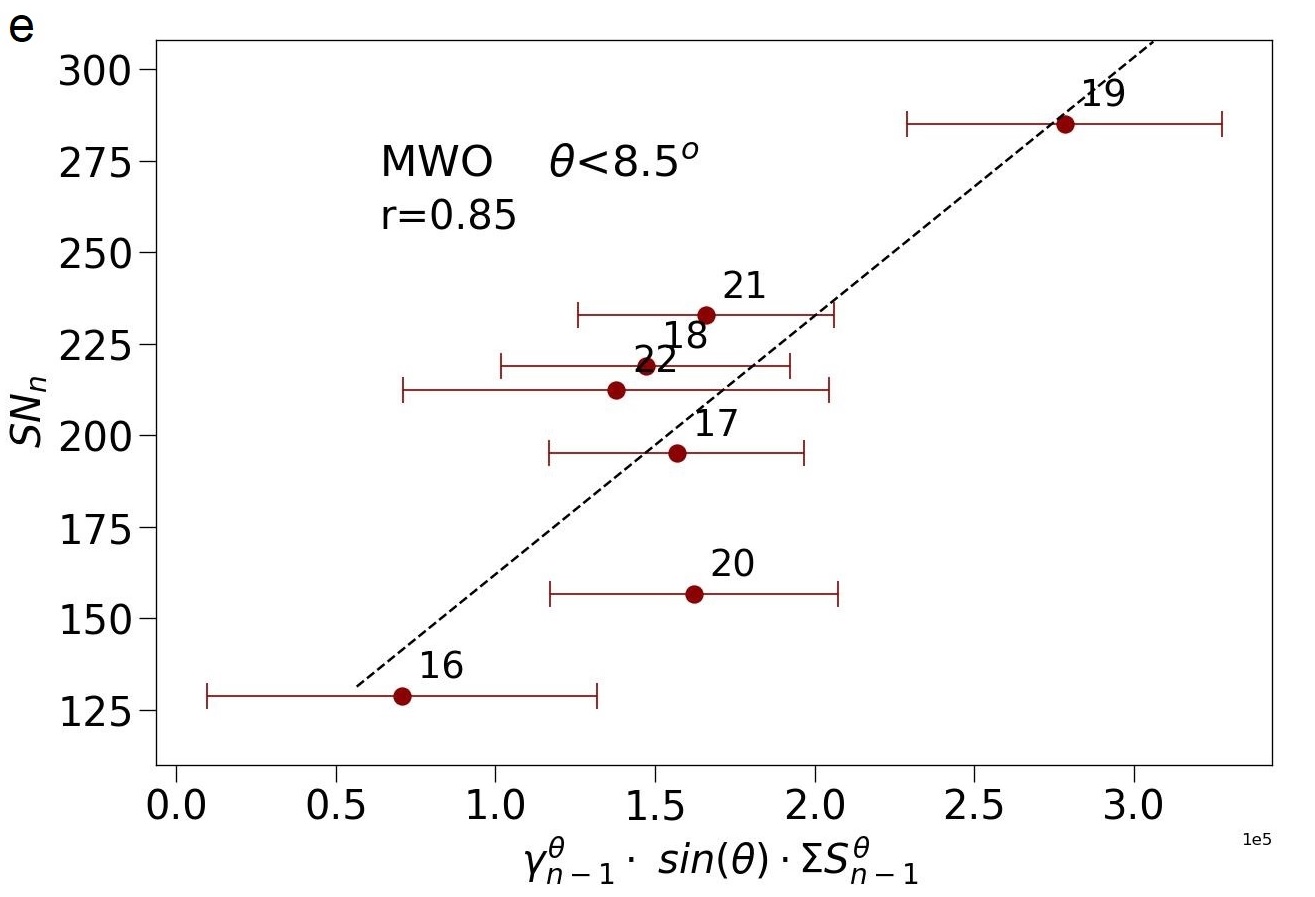}
               \hspace*{0.018\textwidth}
               \includegraphics[width=0.48\textwidth,clip=]{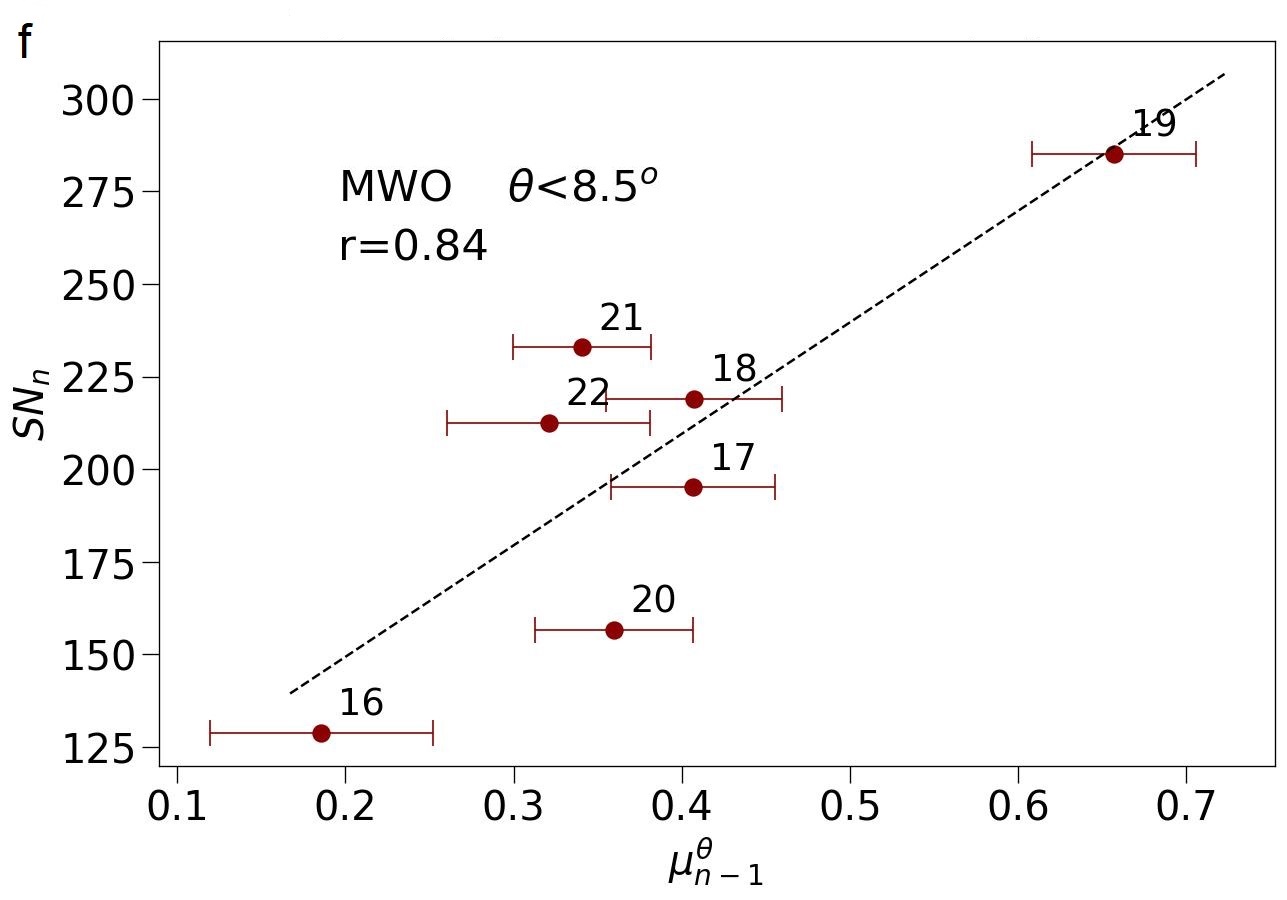}               
\caption{Relationship between the parameters of bipoles at latitudes $\theta<\theta_{cut}$ and the amplitude of the next $SN_{n}$ activity cycle: a) for MWOMF data and the parameter $\overline{\sin\theta\cdot\gamma^{\theta}}_{n-1}\Sigma S_{n-1}^{\theta}$ for $\theta<\theta_{cut}=14.2^o$; b) for MWOMF data and the slope parameter $\mu$ for $\theta_{cut}=14.2^o$; c)  same as (a) for KK data and $\theta_{cut}=9.4^o$; d) same as (b) for KK data and $\theta_{cut}=9.4^o$; e) same as (a) for MWO data for $\theta_{cut}=8.5^o$; f) same as (b) for MWO data for $\theta_{cut}=8.5^o$. The dashed line shows the linear approximation.}
\label{fig:fig3}
\end{figure}

\subsubsection{Low-Latitude Bipoles and the Strength of the Next Activity Cycle}

Let us apply this approach to the selection of prognostic indices based on the active-region tilt angle data. Consider the tilt angle parameters of the active region $\overline{\gamma}$ and $\mu$ for AR near the equator.

To select low-latitude active regions, we introduce the sunspot cutoff latitude $\theta_{cut}$.   We will consider AR with latitude $\theta<\theta_{cut}$.   To determine the optimal latitude $\theta_{cut}$. The optimal angle $\theta_{cut}$, at which the correlation with the amplitude of the next cycle by the sunspot index ($SN$) was highest, was selected for each data set by fitting.
We used maximum $SN$ values  of solar cycles 15\,–\,25  from the source of International Sunspot Numbers \citep{Clette2016}.

To study the relationship between low-latitude bipoles in the previous cycle $n-1$ and the amplitude of the next activity cycle $SN_{n}$, we consider the average values of the tilt angles $\overline{\gamma}$ and the latitude slope parameter $\mu$. Table \ref{T2} presents the values for MWOMF, KK and MWO data. 
We calculated the average values of $\overline{\gamma}$ taking into account the weight by the AR area. We can expect that the amplitude of the next activity cycle $SN_{n}$ also depends on the sum of the AR area at these latitudes. For this, we calculated the area parameter $\Sigma S^{\theta}_{n-1}$, where the summation is over all ARs with latitude $\theta<\theta_{cut}$. The latitude $\theta_{cut}$ for each type of data was found by fitting, but it was preserved for the entire time interval in each data set. For the MWOMF data, we selected $\theta_{cut}=14.2^o$. For the KK data, the optimal value is $\theta_{cut}=9.4^o$, for MWO $\theta_{cut}=8.5^o$.

The relationship between $\overline{\gamma}_{n-1}$  and the amplitude of the next cycle $SN_{n}$ is found to increase if the parameter $\overline{\sin\theta \cdot\gamma}_{n-1}$ is used. Table \ref{T2} presents the values of $\overline{\gamma^{\theta}}$, $\overline{\sin\theta\cdot\gamma^{\theta}}$, $\Sigma S^{\theta}$, as well as the area-weighted parameter $\mu$ and the precision of their determination for activity cycles calculated for MWOMF, KK, MWO data.

The MWOMF data set does not include the first 5 years of cycle 15. The data $\overline{\gamma}$ and $\mu$ for MWOMF cycle 15 are given in Table \ref{T2}.
These values differ significantly from the average values. Probably, the method of measuring the magnetic fields of sunspots in cycle 15 was different from the method in other cycles. The sketches for this period show a relatively small number of sunspots in groups. Therefore, for cycle 15, we took the inclination angles between the two largest spots of different polarity in the group. For other cycles, we used the coordinates of sunspot clusters of different polarity in groups to determine $\gamma$. In Figures 3a and 3b, the measurement cycle for cycle 15 is shown in a different color.

Figure 3a shows the diagram of the relationship between the parameter $\sin\theta\cdot\overline{\gamma^{\theta}}\Sigma S^{\theta}$ and the amplitude of the next $SN_{n}$ activity cycle. If we exclude the incomplete cycle 15, then the forecast of cycles for cycles 16\,-\,25 has a correlation with the amplitude of $SN_{n}$ activity $r=0.96$. Note that without taking into account the sine of the latitude between the parameter $\overline{\gamma^{\theta}}\Sigma S^{\theta}$ the correlation is also quite significant $r=0.90$.  Regressions with the parameter $\overline{\sin\theta\cdot\gamma^{\theta}}\Sigma S^{\theta}$ and the amplitude of the next activity cycle for the KK and MWO data are presented in Figures 3c, 3d. The correlation for the MWO data is $r=0.85$, and for the KK data $r=0.93$. Without taking into account $sin\theta$, the correlation for MWO $r=0.79$ and for KK $r=0.88$.

Another prognostic index is the parameter $\mu^{\theta}$ calculated from bipoles with latitude $\theta<\theta_{cut}$. Figures 3b, 3d, 3f show the regressions for the parameter $\mu^{\theta}_{n-1}$  and $SN_{n}$  for the MWOMF, KK and MWO data, respectively. The parameter $\mu^{\theta}$ in the activity cycles was calculated taking into account the area weight. The values of the parameter $\mu^{\theta}$, as well as the accuracy of its determination $\sigma$ are presented in Table \ref{T2}. For the MWOMF data for cycles 15\,-\,24, the forecast with the amplitude of the next cycle was $r=0.90$. For the KK data for cycles $r=0.94$, and for MWO $r=0.84$.

\begin{figure}[ht!]
\centerline{\includegraphics[width=0.6\textwidth,clip=]{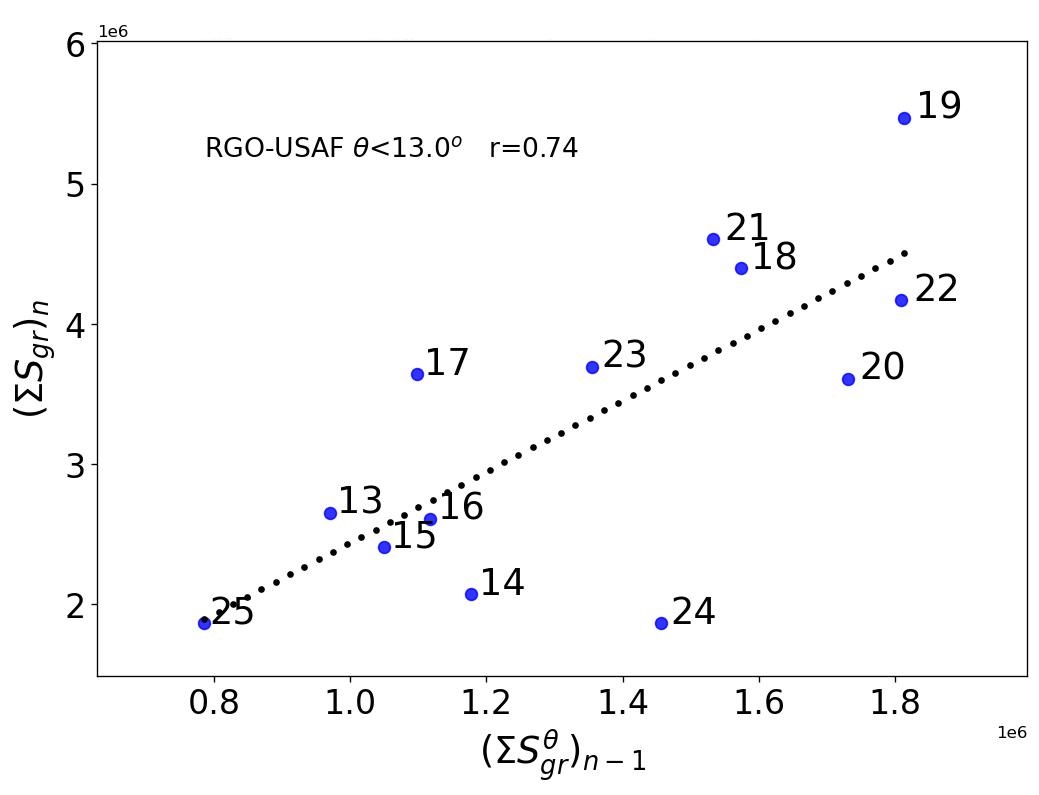}
                 } 
\caption{Relationship between low-latitude activity and the sum of sunspot areas of the next cycle for AR $\theta<13.0^o$, based on RGO-USAF data. The doted line shows the linear approximation.}
\label{fig:fig4}
\end{figure}

\section{Discussion}

In this paper, it was found that the tilt angles of the equatorial active regions have a good correlation with the next activity cycle. This means that for the next activity cycle, not all ARs are important but low-latitude active regions. 
 This hypothesis is consistent with dynamo models that include the influence of meridional circulation on the solar surface \citep{Wang1991}. 
 
 \cite{Dasi-Espuig.etal2010}, to search for the relationship between bipoles and the strength of the next activity cycle for tilt angles $\overline\gamma$, took into account the dependence on latitude $\theta$ by multiplying the exponential function of latitude by tilt angles $e^{-|\theta|/\theta_{o}}$ and weighting by area $S$. They found monthly values of the parameter $\overline\gamma(S,\theta)$ Then they plotted the product of the average monthly sunspot areas $\overline S\overline\gamma(S,\theta)$  against time. The authors then compared the amplitudes of this function with the amplitude of the activity cycle in subsequent cycles. For the characteristic latitude $\theta_{0}=10^{o}$ for the KK and MWO data, they obtained a fairly high correlation $r\approx 0.79$.

At the same time, there are contradictions in the method \citep{Dasi-Espuig.etal2010}, since the properties of predominantly low-latitude bipoles $\overline S \overline\gamma(S,\theta)$ were multiplied by the areas of all sunspots $S$, calculated without the weighting function of latitude. If we take into account the exponential weighting function of latitude also for the areas of sunspots $\overline S$, then we cannot achieve a significant correlation. Note that if we use the areas of sunspots measured in the KK and MWO data and use the monthly values of the parameter $\Sigma S \gamma e^{-|\theta|/\theta_{o}}$, then for $\theta_{0}=13.5^{o}$ the correlation for the KK and MWO data is higher than $r\approx0.85$.

In this analysis we applied a different filtering using the heliographic cutoff latitude $\theta_{cut}$ for the MWOMF, KK and MWO data. The optimal values of $\theta_{cut}$ were different for different data sets. For MWOMF $\theta_{cut}=14.2^o$, for KK   $\theta_{cut}=9.4^o$, for MWO $\theta_{cut}=8.5^o$. The correlation coefficient $r>0.9$ for the MWOMF and KK data (Figures 3a, 3b, 3c and 3d) and $r>0.84$ for MWO data (Figures 3e, 3f).

One of the reasons why different $\theta_{cut}$ are obtained may be different methods of determining areas and coordinates. For the MWOMF data, we combined the bipoles found from magnetic field measurements with solar groups from the RGO-USAF set measured in “white” light. In the KK and MWO data, it was probably not sunspots at the photosphere-penumbra boundary that were measured but the areas and coordinates of sunspot cores. This is evidenced by the relatively small areas in these data series.

It is possible that the connections found between are a consequence of the general connection between low-latitude activity and the strength of the next cycle. It was previously found \citep{Tlatov.etal2024b} that the indices of low-latitude sunspot activity have a good correlation with the amplitude of the next activity cycle.  We use the approach proposed above to search for such connections. Let us consider the index $\Sigma \sin\theta\cdot S^{\theta}$, the sum of sunspot areas at latitudes below a certain value $\theta<\theta_{cut}$. This index does not contain information about the tilt angle AR. Figure \ref{fig:fig4} shows the dependence according to RGO-USAF data for the 12\,-\,24 activity cycles. For this data set $\theta_{cut}=13.0^o$. The correlation is r=0.80. However, using the tilt angle information, the correlation is higher (Figures \ref{fig:fig3}). Also in Figures 3b, 3d and 3f, we use the $\mu$ parameter, which does not sum over the AR area, although the area was used to determine $\mu$ with area weighting. Therefore, it can be concluded that for the next cycle activity, not only the AR area data at low latitudes are important, but also the AR tilt angle information.

This analysis showed that to predict solar cycles, it is necessary to know the properties of magnetic bipoles. To do this, it is necessary to measure the properties of individual sunspots and not just the average parameters of sunspot groups, as in the RGO-USAF data. Currently, the parameters of individual sunspots and pores in groups are measured only at the Kislovodsk Mountian Astronomical Station (KMAS) (\url{http://93.180.26.198:8000/web/grp/}).

The relative share of the total AR area at latitude $\theta<\theta_{cut}=13.0^o$ is $39\%$ of the area of all sunspots (Figure \ref{fig:fig4}). We can assume that sunspots with latitude $\theta>\theta_{cut}$ do not participate in the creation of initial fields for the next cycle. In the dynamo transport models, the main mechanism for generating the magnetic field from initial dipole-type fields is differential rotation. A new activity cycle requires time for the formation of the dipole field, its transfer to the generation zone near the base of the convective zone, and time for the generation and transfer of the toroidal field of the new cycle.  The new activity cycle is not limited to the appearance of sunspots. An extended activity cycle begins several years before the appearance of the first sunspots \citep{McIntosh.eta2022, Martin2023}. In the 24th activity cycle, low-latitude ARs appear in 2013\,-\,2014, and fields of the new cycle are observed at mid-latitudes already in 2016 \citep{Tlatov2025}.  Poloidal magnetic fields from the oppositely polar parts of high-latitude ARs are mutually compensated when transferred to the poles (Figure 2b). Transfer of uncompensated fields from low-latitude ARs to the poles takes about 2 years. This leads to the fact that in traditional transport models there is a time deficit.  Indeed, low-latitude regions that appeared in 2013 cannot lead to fields of a new cycle in 2016. Since at least 4 years are needed to transport fields to the poles and back. This problem can be solved using the hypothesis of the formation of a near-surface azimuthal magnetic field that participates in the formation of a new activity cycle \citep{Tlatov2023}.  In this scheme, the near-surface azimuthal magnetic field of a new cycle is formed already with the first sunspots appearing at high latitudes. But low-latitude ARs have a special role for a new solar cycle. For such bipoles, whose magnetic parts are on different sides of the equator, differential rotation forms an azimuthal magnetic field, which can be strengthened by differential rotation for a long time and is not carried away by meridional circulation. Azimuthal near-surface magnetic fields from low-latitude bipoles serve as initial fields for sunspots of a new cycle.

\begin{acknowledgments}
This study includes data from the synoptic program at the 150-Foot Solar Tower of the Mt. Wilson Observatory Digitization of these data was carried out by K.A. Tlatova.  The author acknowledges the work of all the people who contribute to and maintain the GPR-USAF, MWO and KK sunspot databases. We used maximum values of solar cycles 15–25 from the Source of International Sunspot Numbers: WDC-SILSO, Royal Observatory of Belgium, Brussels. 
The author also acknowledge the financial support of the Ministry of Science and Higher Education of the Russian Federation, grant number 075-03-2025-420/4.
\end{acknowledgments}

\bibliography{spot_tilt.bbl}{}
\bibliographystyle{aasjournal}

\end{document}